\documentclass[12pt]{article}
 \pdfoutput=1
\usepackage{amssymb}
\usepackage{amsmath}
\usepackage{amstext}
\usepackage{graphicx,epsfig}
\usepackage{epsfig}
\usepackage{verbatim} 
\usepackage{fancyhdr}
\usepackage{fancybox}
\usepackage{color}
\usepackage{enumitem}
\usepackage{subfigure}
\usepackage{parskip}
\usepackage{cite}

\newcommand{\Comment}[1]{{}}
\definecolor{MyDarkBlue}{rgb}{0.15,0.15,0.45}
\definecolor{reddish}{rgb}{0.65, 0.2, 0.2}
\usepackage[linktocpage=true]{hyperref}
\hypersetup{
colorlinks=true,
citecolor=MyDarkBlue,
linkcolor=reddish,
urlcolor=MyDarkBlue,
pdfauthor={},
pdftitle={},
pdfsubject={}
}

\newcommand{\be}{\begin{equation}}
\newcommand{\ee}{\end{equation}}
\newcommand{\bea}{\begin{eqnarray}}
\newcommand{\eea}{\end{eqnarray}}
\newcommand{\beas}{\begin{eqnarray*}}
\newcommand{\eeas}{\end{eqnarray*}}
\newcommand{\nn}{\nonumber}

\newcommand{\half}{\frac{1}{2}}

\numberwithin{equation}{section}

\begin{document}

\renewcommand{\thefootnote}{\fnsymbol{footnote}}
~
\vspace{.75truecm}

\begin{center}
{\Large \bf{No-go for Partially Massless Spin-2 Yang--Mills}}
\end{center} 
 \vspace{1truecm}
\thispagestyle{empty} 

\centerline{
{\large Sebastian Garcia-Saenz,$^{\rm a,}$\footnote{\href{mailto:sgsaenz@phys.columbia.edu}{\tt sgsaenz@phys.columbia.edu}} 
Kurt Hinterbichler,$^{\rm b,}$\footnote{\href{mailto:khinterbichler@perimeterinstitute.ca}{\tt khinterbichler@perimeterinstitute.ca}} 
Austin Joyce,${}^{\rm c,}$\footnote{\href{mailto:ajoy@uchicago.edu}{\tt ajoy@uchicago.edu}
}}}
\centerline{
{\large  Ermis Mitsou${}^{\rm a,}$\footnote{\href{mailto:ermis.mitsou@columbia.edu}{\tt ermis.mitsou@columbia.edu}}
and Rachel A. Rosen${}^{\rm a,b,}$\footnote{\href{mailto:rar2172@columbia.edu}{\tt rar2172@columbia.edu}
}}}

\vspace{.75cm}

\centerline{{\it ${}^{\rm a}$Physics Department and Institute for Strings, Cosmology, and Astroparticle Physics,}}
 \centerline{{\it Columbia University, New York, NY 10027, USA}}

\vspace{.5cm}

\centerline{{\it ${}^{\rm b}$Perimeter Institute for Theoretical Physics,}}
\centerline{{\it 31 Caroline St. N, Waterloo, Ontario, Canada, N2L 2Y5 }}
 
 \vspace{.5cm}

\centerline{\it ${}^{\rm c}$Enrico Fermi Institute and Kavli Institute for Cosmological Physics,}
\centerline{\it University of Chicago, Chicago, IL 60637}

 \vspace{.5cm}

\begin{abstract} 
\noindent
There are various no-go results forbidding self-interactions for a single partially massless spin-2 field.  Given the photon-like structure of the linear partially massless field, it is natural to ask whether a multiplet of such fields can interact under an internal Yang--Mills like extension of the partially massless symmetry.  We give two arguments that such a partially massless Yang--Mills theory does not exist.  The first is that there is no Yang--Mills like non-abelian deformation of the partially massless symmetry, and the second is that  cubic vertices with the appropriate structure constants do not exist.

\end{abstract}

\newpage
\renewcommand*{\thefootnote}{\arabic{footnote}}



\newpage
\setcounter{footnote}{0}

\section{Introduction}
\parskip=2pt
\normalsize

Massive higher spin fields on de Sitter space possess gauge symmetries at certain values of their masses \cite{Deser:1983tm,Deser:1983mm,Higuchi:1986py,Brink:2000ag,Deser:2001pe,Deser:2001us,Deser:2001wx,Deser:2001xr,Zinoviev:2001dt,Garidi:2003ys,Skvortsov:2006at,Skvortsov:2009zu}.  The first example of a field with more than one distinguished mass value is a massive spin-2, $h_{\mu\nu}$, on a $D$ dimensional de Sitter space of radius $H^{-1}$, which has the action
 \begin{align}
 \nn S=\int {\rm d}^Dx\sqrt{-g}\left[ -{1\over 2}\nabla_\lambda h_{\mu\nu} \nabla^\lambda h^{\mu\nu}+\nabla_\lambda h_{\mu\nu} \nabla^\nu h^{\mu\lambda}-\nabla_\mu h\nabla_\nu h^{\mu\nu}+\half \nabla_\mu h\nabla^\mu h\right. \\
\left.+\left(D-1\right)H^2\left( h^{\mu\nu}h_{\mu\nu}-\half h^2\right)-\frac{1}{2}m^2(h_{\mu\nu}h^{\mu\nu}-h^2)\right]. \label{curvedmassivelin}
\end{align}
When the mass takes the value
\be 
m^2=\left(D-2\right)H^2\, ,\label{masstuning}
\ee
the theory develops a scalar gauge symmetry
\be 
\delta h_{\mu\nu}= \left(\nabla_\mu\nabla_\nu+H^2 g_{\mu\nu}\right)\phi\, ,\label{gaugesym}
\ee
where $\phi(x)$ is a scalar gauge parameter.  This is the partially massless (PM) graviton.  There is of course another distinguished value of the mass, $m=0$, corresponding to the ordinary massless graviton, which is invariant under linear diffeomorphism invariance.  These are the only two values of the mass of a spin-2 for which a gauge symmetry appears. In four dimensions, a generic massive spin-2 field propagates five degrees of freedom, a massless spin-2 field propagates two degrees of freedom, and a partially massless spin-2 lies in-between, propagating 4 degrees of freedom.\footnote{More generally, in $D$ dimensions, the PM graviton propagates $N_{\rm dof} = \tfrac{(D+1)(D-2)}{2}-1$ degrees of freedom.}

This partially massless theory has been of interest as a possible theory of gravity because the symmetry-enforced relation \eqref{masstuning} links the value of the cosmological constant to the graviton mass.   A small graviton mass is in turn technically natural due to the enhanced diffeomorphism invariance of general relativity at the value $m=0$~\cite{deRham:2012ew,deRham:2013qqa}. This offers a tantalizing possible avenue towards solving the cosmological constant problem~\cite{deRham:2013wv}.  Unfortunately, there are obstructions to realizing a complete two-derivative non-linear theory that maintains the gauge symmetry and propagates the same number of degrees of freedom as the linear theory~\cite{Zinoviev:2006im,Deser:2012qg,deRham:2013wv,Joung:2014aba,Zinoviev:2014zka,Garcia-Saenz:2014cwa}.

On the other hand, the linear partially massless spin-2 theory shares many properties with ordinary Maxwell electrodynamics, including null propagation in four dimensions~\cite{Deser:2001xr}, a scalar gauge symmetry~\eqref{gaugesym}, a one-derivative gauge invariant field strength tensor \eqref{fieldstrengthfd}, duality invariance~\cite{Deser:2013xb,Hinterbichler:2014xga} and monopole solutions~\cite{Hinterbichler:2015nua}.  Another property shared by the partially massless graviton and the photon is the aforementioned difficulty with constructing self interactions; there are no non-trivial two-derivative self interaction terms for a single photon.  The only self-interactions we can write in general dimensions are powers of the Maxwell field strength and its derivatives.  These Born--Infeld~\cite{Born:1934gh} or Euler-Heisenberg~\cite{Heisenberg:1935qt} like interactions are invariant under the {linear} gauge symmetry and have as many derivatives as fields. Similarly, we can take arbitrary powers of the PM field strength and construct a theory which is nonlinear but invariant under the linear gauge symmetry.\footnote{Equations of motion of this sort which are ghost-free and invariant under duality are constructed in \cite{Cherney:2015jxp}.} However, these higher derivative interactions are not relevant at low energies, and we would like to construct interactions with {fewer} derivatives which may deform the gauge symmetry in a nontrivial manner.\footnote{In odd dimensions, there exist Chern--Simons interactions which have fewer derivatives per field than terms constructed from powers of the field strength, but which are nevertheless invariant under linear gauge transformations \cite{Deser:1981wh}.}

If we have a multiplet of photons $A_\mu^a$, labelled by some color index $a$ (raised and lowered with $\delta_{ab}$), we know there is a way to achieve this through Yang--Mills theory \cite{Yang:1954ek},
\begin{align}
{\cal L}
=-\frac{1}{4}(\partial_\mu A_{\nu a}-\partial_{\nu}
A_{\mu a})(\partial^\mu A^{\nu a}-\partial^\nu A^{\mu a})-gf_{abc}^{\ \ }\partial_\mu A_{\nu }^aA^{\mu b}A^{\nu c}-\frac{g^2}{4}f_{abc}f_{de}^{\ \ a} A_\mu^b A_\nu^c A^{\mu d}A^{\nu e}. \label{YMlag}
\end{align}
Here $g$ is the coupling constant and $f_{abc}$ is some order one structure tensor. Consistency of the gauge symmetry at cubic order requires the structure tensor to be fully antisymmetric, and consistency at quartic order demands that it satisfy the Jacobi identity $f^{\ \ d}_{ac}f^{\ \ c}_{be}+f^{\ \ d}_{bc}f^{\ \ c}_{ea}+f^{\ \ d}_{ec}f^{\ \ c}_{ab}=0$.   The gauge symmetry is deformed from the abelian $\delta A_\mu=\partial_\mu \alpha$ into the non-abelian
\be 
\delta A_\mu^a=\partial_\mu\alpha^a+f_{bc}^{\ \ a}A_\mu^b\alpha^c.
\ee
Due to the total antisymmetry of $f_{abc}$, at least 3 photons are required to write a Yang--Mills interaction\footnote{See \cite{Deser:1963zzc} for more constraints on the possible interactions of spin 1.} (in which case the gauge symmetries close to form the $su(2)$ algebra).  

Given the similarities between the partially massless spin-2 and the photon, a natural question to ask is whether the self-interaction difficulties of the partially massless spin-2 can be obviated in the same way -- by extending to a multiplet of fields.  This is the question we address in this paper: {\it does there exist a Yang--Mills like theory for an interacting multiplet, $h_{\mu\nu}^a$, of partially massless spin-2 fields?}\footnote{Such Yang--Mills type theories for {\it massless} gravitons do not exist~\cite{Boulanger:2000rq}, but this no-go result does not immediately rule out PM Yang--Mills, as the form of the gauge symmetry is entirely different.}

The ideal form of such a theory would be a two-derivative action with interactions governed by an antisymmetric coupling $f_{abc}$, schematically of the form
\be {\cal L}\sim \left(\nabla^2 h^2+{f\over M} (\nabla^2h^3+H^2 h^3) +{f^2\over M^2} (\nabla^2 h^4+H^2 h^4) +\cdots \right)\, .\label{YMpa}\ee
Here $M$ is some mass scale (in gravitational applications it would be the Planck mass) suppressing the powers of the field (using, for illustration, canonical normalization appropriate to the $D=4$ case).
In the flat limit, $H\to0$, the degrees of freedom of a partially massless graviton reduce to those of a massless graviton and a massless vector.  This can be seen at the level of the action by introducing the vector through a St\"uckelberg replacement $h_{\mu\nu}\mapsto h_{\mu\nu}+ {1\over H}\left( \nabla_\mu A_{\nu}+ \nabla_\nu A_{\mu}\right)$ and then taking the $H\to 0$ limit \cite{deRham:2013wv}, in which the partially massless symmetry \eqref{gaugesym} becomes the $U(1)$ symmetry of the vector.\footnote{The vector kinetic term has the wrong sign on anti-de Sitter space, and the correct sign for the theory in de Sitter space, reflecting the fact that the theory is unitary only in the de Sitter case.} If we were to take this limit in our hypothetical partially massless Yang--Mills lagrangian~\eqref{YMpa}, ideally it would reduce to the spin-1 Yang--Mills theory~\eqref{YMlag}, in addition to a decoupled massless graviton.  Because the St\"uckelberg replacement always brings a derivative along with each power of $A$, there would have to be a cancellation among the highest derivatives of \eqref{YMpa} in order to yield the action \eqref{YMlag}, whose interactions have fewer derivatives than fields.  Such cancellations would leave powers of $H$, resulting in a gauge coupling $g\sim {H\over M}$.

In the case of electromagnetism, there is a $U(1)$ invariant field strength $F_{\mu\nu}=\partial_\mu A_\nu-\partial_\nu A_\mu$, and the lagrangian can be written as its square: ${\cal L}=-{1\over 4}F_{\mu\nu}^2$.  Yang--Mills theory can be arrived at by first finding a non-abelian version of the field strength which contains quadratic powers of the field and transforms covariantly under the gauge symmetry:
\be
F_{\mu\nu}^a=\partial_\mu A_\nu^a-\partial_\nu
A_\mu^a+f_{bc}^{\ \ a}A_{\mu}^bA_\nu^c,~~~~~~~~~~~~~~ \delta F_{\mu\nu}^a=f_{bc}^{\ \ a}\alpha^b F_{\mu\nu}^c.
\ee
The full Yang--Mills lagrangian is then easily written as ${\cal L}=-{1\over 4}(F^a_{\mu\nu})^2$.  The partially massless theory also has a gauge invariant field strength \cite{Deser:2006zx},
\be F_{\mu\nu\rho}=\nabla_\mu h_{\nu\rho}-\nabla_\nu h_{\mu\rho},\label{fieldstrengthfd}\ee
and the partially massless action can be written as a particular combination of its squares,
\be 
S=-\frac{1}{4} \int{\rm d}^Dx\sqrt{-g}\left(F^{\lambda\mu\nu}F_{\lambda\mu\nu}-2F^{\lambda\mu}_{\ \ \ \mu} F_{\lambda\nu}^{\ \ \nu}\right). \ee
A natural first attempt is to try to find a non-abelian version of this curvature,
\be F_{\mu\nu\rho}^a=\nabla_\mu h_{\nu\rho}^a-\nabla_\nu h_{\mu\rho}^a+{\cal O}\left(h^2\right),\label{ymfa}\ee
and then to write the full non-linear action as
\be {\cal L}=-\frac{1}{4}\int{\rm d}^Dx \sqrt{-g}  \left(F^{\lambda\mu\nu}_aF_{\lambda\mu\nu}^a-2F^{\lambda\mu}_{a \ \ \ \mu } F_{\lambda\nu}^{a \ \ \nu }\right). \ee
However, by trying all possible combinations of $h^2$ terms in \eqref{ymfa}, it is not hard to become convinced that it is impossible to construct such a non-abelian curvature in this case.

In what follows we will provide two general arguments that, indeed, no theory that we might reasonably call a partially massless Yang--Mills like theory is possible.  First, we use the requirement that two gauge symmetries must commute to another gauge symmetry to show that the structure of the gauge algebra for a multiplet of partially massless fields must be abelian, thereby ruling out the non-commutativity essential to a Yang--Mills like theory.  Second, by studying dual CFT correlation functions, we show that there is no possible non-trivial antisymmetric cubic vertex through which a multiplet of partially massless fields might interact.  

\section{Closure argument\label{closuresect}}

A powerful tool in the search for nonlinear deformations of gauge symmetries is the {\it closure condition} (sometimes called the admissibility condition \cite{Konshtein:1988yg} in certain cases).   It is the requirement that for any two gauge parameters, $\phi^a$ and $\psi^a$, the equation
\begin{equation} \label{eq:closure_general}
\left[\delta_{\phi},\delta_{\psi}\right]h_{\alpha\beta}^{a}=\delta_{\chi}h_{\alpha\beta}^{a}+{\rm on\ shell\ trivial}\,,
\end{equation}
holds for some function $\chi^a$. In other words, the gauge symmetries must form an algebra up to on-shell trivial symmetries, as this is a necessary condition for the gauge orbits in field space to be integrable, { i.e.}, that the infinitesimal transformation derives from a finite one.

The closure condition was applied by Wald in \cite{Wald:1986bj} to obtain Yang--Mills theory and general relativity as the unique nonlinear theories of multiple spin-1 fields and of a single spin-2 field, respectively, assuming the interactions do not contain more than two derivatives. This has been extended to a collection of spin-2 fields in \cite{Cutler:1986dv,Wald:1986dw}, and more recently to theories with a single partially massless spin-2 field \cite{Joung:2014aba,Garcia-Saenz:2014cwa}. The goal of this section is to analyze the case of multiple PM spin-2 fields by looking at the lowest order part of the closure condition (\ref{eq:closure_general}). We will see that this is already enough to establish that the free PM gauge symmetry cannot be extended to a field-dependent one of the Yang--Mills type.  This analysis is an extension of those of~\cite{Wald:1986bj,Garcia-Saenz:2014cwa}.

The power of the closure condition lies in its generality, as one does not need to make any assumptions regarding the action of the nonlinear theory. In particular, one does not have to assume that interactions appear at a certain order in powers of the fields, nor must one restrict the number of derivatives in the lagrangian. Our only assumptions are  that the gauge transformation reduces to the usual free PM symmetry at zeroth order in powers of the field,
\begin{equation}
\delta^{(0)}h_{\mu\nu}^{a}=\left({\nabla}_{\mu}{\nabla}_{\nu}+H^2{g}_{\mu\nu}\right)\phi^{a}\,,
\end{equation}
and that the full transformation only involves terms with up to two derivatives.

The most general two-derivative extension of the partially massless gauge symmetry to nonlinear order may be written as
\begin{equation} 
\label{eq:gauge_sym1}
\delta_{\phi}h_{\alpha\beta}^{a}=B^{\mu\nu\phantom{\alpha\beta}a}_{\phantom{\mu\nu}\alpha\beta\phantom{a}b}\left({\nabla}_{\mu}{\nabla}_{\nu}\phi^{b}+D^{\lambda\phantom{\mu\nu}b}_{\phantom{\lambda}\mu\nu\phantom{b}c}\,{\nabla}_{\lambda}\phi^{c}+C_{\mu\nu\phantom{b}c}^{\phantom{\mu\nu}b}\,\phi^{c}\right)\,.
\end{equation}
Here, the tensor $B$ is assumed to contain only powers of $h^a$ with no derivatives, $D$ must be at most linear in ${\nabla}h^a$, and $C$ may have terms linear in ${\nabla}{\nabla}h^a$, quadratic in ${\nabla}h^a$, as well as arbitrary powers of $h^a$.  As in \cite{Garcia-Saenz:2014cwa}, we may take the $\mu\nu$ index pairs in \eqref{eq:gauge_sym1} to be symmetric without loss of generality.

We will expand each of the tensors $B, D, C$ in powers of the field, $h$, and demand that the closure condition be satisfied order-by-order. This will fix the form of these tensors.  At lowest order, where the tensors are field-independent, we require the gauge symmetry~\eqref{eq:gauge_sym1} to reduce to the free PM transformation.  This fixes
\begin{equation}
B^{(0)\mu\nu\phantom{\alpha\beta}a}_{\phantom{(0)\mu\nu}\alpha\beta\phantom{a}b}=\delta^{\mu}_{(\alpha}\delta^{\nu}_{\beta)}\delta^a_b\,,\qquad D^{(0)\lambda\phantom{\mu\nu}b}_{\phantom{(0)\lambda}\mu\nu\phantom{b}c}=0\,,\qquad C_{\phantom{(0)}\mu\nu\phantom{b}c}^{(0)\phantom{\mu\nu}b}=H^2{g}_{\mu\nu}\delta^b_c\,.
\end{equation}
Here, the superscript $(n)$ denotes $n$ powers of the fields. At the next order in the expansion, the most general expression we can write is
\begin{equation} \label{eq:order1tensors}
\begin{split}
B^{(1)\mu\nu\phantom{\alpha\beta}a}_{\phantom{(1)\mu\nu}\alpha\beta\phantom{a}b}&= b_{1\phantom{a}bc}^{\phantom{1}a}\,{g}^{\mu\nu}h_{\alpha\beta}^{c}+b_{2\phantom{a}bc}^{\phantom{2}a}\,{g}_{\alpha\beta}h^{c\,\mu\nu}+b_{3\phantom{a}bc}^{\phantom{3}a}\,{g}^{\mu\nu}{g}_{\alpha\beta}h^c+b_{4\phantom{a}bc}^{\phantom{4}a}\,\delta^{\mu}_{(\alpha}\delta^{\nu}_{\beta)}h^c\\
&\quad+b_{5\phantom{a}bc}^{\phantom{5}a}\,\delta^{(\mu}_{(\alpha}h_{\beta)}^{\nu)\,\,c}\,,\\
D^{(1)\lambda\phantom{\alpha\beta}a}_{\phantom{(1)\lambda}\alpha\beta\phantom{a}b}&=d_{1\phantom{a}bc}^{\phantom{1}a}\,{\nabla}^{\lambda}h_{\alpha\beta}^{c}+d_{2\phantom{a}bc}^{\phantom{2}a}\,{\nabla}_{(\alpha}h_{\beta)}^{c\,\,\lambda}+d_{3\phantom{a}bc}^{\phantom{3}a}\,\delta^{\lambda}_{(\alpha}{\nabla}_{\beta)}h^c+d_{4\phantom{a}bc}^{\phantom{4}a}\,\delta^{\lambda}_{(\alpha}{\nabla}^{\sigma}h_{\beta)\sigma}^{c}\\
&\quad+d_{5\phantom{a}bc}^{\phantom{5}a}\,{g}_{\alpha\beta}{\nabla}^{\lambda}h^c+d_{6\phantom{a}bc}^{\phantom{6}a}\,{g}_{\alpha\beta}{\nabla}_{\sigma}h^{c\,\lambda\sigma}\,,\\
C_{\phantom{(1)}\alpha\beta\phantom{a}b}^{(1)\phantom{\alpha\beta}a}&= c_{1\phantom{a}bc}^{\phantom{1}a}\,{\nabla}_{\alpha}{\nabla}_{\beta}h^c+c_{2\phantom{a}bc}^{\phantom{2}a}\,{\nabla}_{\sigma}{\nabla}_{(\alpha}h_{\beta)}^{c\,\,\sigma}+c_{3\phantom{a}bc}^{\phantom{3}a}\,\Box h_{\alpha\beta}^{c}+c_{4\phantom{a}bc}^{\phantom{4}a}\,{g}_{\alpha\beta}\Box h^c\\
&\quad+c_{5\phantom{a}bc}^{\phantom{5}a}\,{g}_{\alpha\beta}{\nabla}_{\lambda}{\nabla}_{\sigma}h^{c\,\lambda\sigma}+c_{6\phantom{a}bc}^{\phantom{6}a}H^2\,h_{\alpha\beta}^{c}+c_{7\phantom{a}bc}^{\phantom{7}a}H^2\,{g}_{\alpha\beta}h^c\,,
\end{split}
\end{equation}
where the constant tensors $b_{n~bc}^{~a}$, $d_{n~bc}^{~a}$ and $c_{n~bc}^{~a}$ will be determined by the closure condition~\eqref{eq:closure_general}.

In writing this, we must be aware that there exists some redundancy in the definition of the tensors $B$, $D$ and $C$ arising from the fact that we are free to perform a redefinition of either the fields $h^a$ or the gauge parameter $\phi^a$. In the first case, we note that a field redefinition of the form
\begin{equation}
h_{\alpha\beta}^{a}\mapsto \tilde{h}_{\alpha\beta}^{a}(h)\,,
\end{equation}
amounts, in the symmetry transformation~\eqref{eq:gauge_sym1}, to changing the $B$ tensor as
\begin{equation}
B^{\mu\nu\phantom{\alpha\beta}a}_{\phantom{\mu\nu}\alpha\beta\phantom{a}b}\mapsto \frac{\partial \tilde{h}_{\alpha\beta}^{a}}{\partial h_{\lambda\sigma}^{c}}\,B^{\mu\nu\phantom{\lambda\sigma}c}_{\phantom{\mu\nu}\lambda\sigma\phantom{c}b} \, , \hspace{1cm} C_{\mu\nu\phantom{a}b}^{\phantom{\mu\nu}a} \mapsto C_{\mu\nu\phantom{a}b}^{\phantom{\mu\nu}a} \, , \hspace{1cm} D^{\lambda\phantom{\mu\nu}a}_{\phantom{\lambda}\mu\nu\phantom{a}b} \mapsto D^{\lambda\phantom{\mu\nu}a}_{\phantom{\lambda}\mu\nu\phantom{a}b} \,.
\end{equation}
In the second case, a redefinition of the gauge parameter 
\begin{equation}
\phi^a\mapsto f^a_{\phantom{a}b}(h)\phi^b \, ,
\end{equation}
with $f^a_{\phantom{a}b}$ being an arbitrary function of the fields $h^a$, can be compensated by shifting the $B$, $D$ and $C$ tensors as
\begin{equation} \label{eq:f_redef}
\begin{split}
B^{\mu\nu\phantom{\alpha\beta}a}_{\phantom{\mu\nu}\alpha\beta\phantom{a}b}&\mapsto B^{\mu\nu\phantom{\alpha\beta}a}_{\phantom{\mu\nu}\alpha\beta\phantom{a}c}\,f^{c}_{\phantom{c}b}\,,\\
D^{\lambda\phantom{\mu\nu}a}_{\phantom{\lambda}\mu\nu\phantom{a}b}&\mapsto (f^{-1})^{a}_{\phantom{a}c}\,D^{\lambda\phantom{\mu\nu}c}_{\phantom{\lambda}\mu\nu\phantom{c}d}\,f^{d}_{\phantom{d}b}+2(f^{-1})^{a}_{\phantom{a}c}\,\delta^{\lambda}_{(\mu}{\nabla}_{\nu)}f^{c}_{\phantom{c}b}\,,\\
C_{\mu\nu\phantom{a}b}^{\phantom{\mu\nu}a}&\mapsto (f^{-1})^{a}_{\phantom{a}c}\,C_{\mu\nu\phantom{c}d}^{\phantom{\mu\nu}c}\,f^{d}_{\phantom{d}b}+(f^{-1})^{a}_{\phantom{a}c}\,{\nabla}_{\mu}{\nabla}_{\nu}f^{c}_{\phantom{c}b}+(f^{-1})^{a}_{\phantom{a}c}\,D^{\lambda\phantom{\mu\nu}c}_{\phantom{\lambda}\mu\nu\phantom{c}d}\,{\nabla}_{\lambda}f^{d}_{\phantom{d}b} \,.
\end{split}
\end{equation}
This can be inferred by observing that the Noether identity that follows from eq.~\eqref{eq:gauge_sym1},
\begin{equation}
{\nabla}_{\mu}{\nabla}_{\nu}\big(B^{\mu\nu\phantom{\alpha\beta}a}_{\phantom{\mu\nu}\alpha\beta\phantom{a}b}\,\mathcal{E}^{\alpha\beta}_a\big)-{\nabla}_{\lambda}\big(B^{\mu\nu\phantom{\alpha\beta}a}_{\phantom{\mu\nu}\alpha\beta\phantom{a}c}D^{\lambda\phantom{\mu\nu}c}_{\phantom{\lambda}\mu\nu\phantom{c}b}\,\mathcal{E}^{\alpha\beta}_a\big)+B^{\mu\nu\phantom{\alpha\beta}a}_{\phantom{\mu\nu}\alpha\beta\phantom{a}c}C_{\mu\nu\phantom{c}b}^{\phantom{\mu\nu}c}\,\mathcal{E}^{\alpha\beta}_a=0\,,
\end{equation}
where $\mathcal{E}^{\alpha\beta}_a\equiv \delta S/\delta h_{\alpha\beta}^a$ is the equation of motion, remains unchanged by the redefinitions in~\eqref{eq:f_redef}, which therefore do not affect the constraints of the putative nonlinear PM theory.

It is useful to fix these redundancies by simplifying the tensor $B^{(1)}$. The most general $\phi^a$ redefinition at this order has $f^a_{\phantom{a}b}=\delta^a_b+\alpha^a_{\phantom{\alpha}bc}h^c$, for some constant $\alpha^a_{\phantom{\alpha}bc}$, under which $B^{(1)}$ changes as
\begin{equation}
B^{(1)\mu\nu\phantom{\alpha\beta}a}_{\phantom{(0)\mu\nu}\alpha\beta\phantom{a}b}\mapsto B^{(1)\mu\nu\phantom{\alpha\beta}a}_{\phantom{(0)\mu\nu}\alpha\beta\phantom{a}b}+\alpha^a_{\phantom{a}bc}\,\delta^{\mu}_{(\alpha}\delta^{\nu}_{\beta)}h^c\,;
\end{equation}
this allows us to set $b_4=0$ in $B^{(1)}$. (Alternatively, we could choose $f^a_{\phantom{a}b}$ to eliminate $d_{3}$ or $c_1$.) Secondly, the most general field redefinition which affects the first order tensors is a quadratic one
\begin{equation}
\tilde{h}_{\alpha\beta}^{a}={h}_{\alpha\beta}^{a}+a_{1\phantom{a}bc}^{\phantom{1}a}\,{g}_{\alpha\beta}h_{\lambda\sigma}^{b}h^{c\,\lambda\sigma}+a_{2\phantom{a}bc}^{\phantom{2}a}\,{g}_{\alpha\beta}h^bh^c+a_{3\phantom{a}bc}^{\phantom{3}a}\,h^bh_{\alpha\beta}^{c}+a_{4\phantom{a}bc}^{\phantom{4}a}\,h_{(\alpha}^{b\,\lambda}h_{\beta)\lambda}^{c}\,,
\end{equation}
which shifts $B^{(1)}$ by
\begin{equation}
\begin{split}
\frac{\partial \tilde{h}_{\alpha\beta}^{a}}{\partial h_{\mu\nu}^{b}}&=a_{3\phantom{a}bc}^{\phantom{3}a}\,{g}^{\mu\nu}h_{\alpha\beta}^{\phantom{\alpha\beta}c}+2a_{1\phantom{a}(bc)}^{\phantom{1}a}\,{g}_{\alpha\beta}h^{c\,\mu\nu}+2a_{2\phantom{a}(bc)}^{\phantom{2}a}\,{g}^{\mu\nu}{g}_{\alpha\beta}h^c\\
&\quad+a_{3\phantom{a}cb}^{\phantom{3}a}\,\delta^{\mu}_{(\alpha}\delta^{\nu}_{\beta)}h^c+2a_{4\phantom{a}(bc)}^{\phantom{4}a}\,\delta^{(\mu}_{(\alpha}h_{\beta)}^{\phantom{\beta)}\nu)c}\,.
\end{split}
\end{equation}
The $a_4$ term can be set to zero by choosing $\alpha^a_{\phantom{a}bc}$ appropriately. We therefore see that we have the freedom to set $b_{1\phantom{a}bc}^{\phantom{I}a}=0$, $b_{2\phantom{a}(bc)}^{\phantom{2}a}=0$, $b_{3\phantom{a}(bc)}^{\phantom{3}a}=0$, and $b_{5\phantom{a}(bc)}^{\phantom{5}a}=0$. Thus we end up with
\begin{equation}
B^{(1)\mu\nu\phantom{\alpha\beta}a}_{\phantom{(1)\mu\nu}\alpha\beta\phantom{a}b}= b_{2\phantom{a}[bc]}^{\phantom{2}a}\,{g}_{\alpha\beta}h^{c\,\mu\nu}+b_{3\phantom{a}[bc]}^{\phantom{3}a}\,{g}^{\mu\nu}{g}_{\alpha\beta}h^c+b_{5\phantom{a}[bc]}^{\phantom{5}a}\,\delta^{(\mu}_{(\alpha}h_{\beta)}^{\phantom{\beta)}\nu)c}\,,
\end{equation}
where the surviving $b$ coefficients are antisymmetric in their lower indices.

Having eliminated all the redundancies in the PM gauge symmetry at first order in the fields, we may now look at the constraints that arise from the closure condition at the lowest nontrivial order:
\begin{equation} \label{eq:closure_order0}
\left(\delta^{(0)}_{\phi}\delta^{(1)}_{\psi}-\delta^{(0)}_{\psi}\delta^{(1)}_{\phi}\right)h_{\alpha\beta}^{a}=\delta^{(0)}_{\chi^{(0)}}h_{\alpha\beta}^{a}\,.
\end{equation}
Writing eq.~\eqref{eq:closure_order0} more explicitly, we find the equation
\begin{equation}
\begin{split}
&\left(\delta^{(0)}_{\phi}B^{(1)\mu\nu\phantom{\alpha\beta}a}_{\phantom{(1)\mu\nu}\alpha\beta\phantom{a}b}\right)\left({\nabla}_{\mu}{\nabla}_{\nu}\psi^b+H^2{g}_{\mu\nu}\psi^b\right)+\left(\delta^{(0)}_{\phi}D^{(1)\lambda\phantom{\alpha\beta}a}_{\phantom{(1)\lambda}\alpha\beta\phantom{a}b}\right){\nabla}_{\lambda}\psi^b\\
&+\left(\delta^{(0)}_{\phi}C_{\phantom{(1)}\alpha\beta\phantom{a}b}^{(1)\phantom{\alpha\beta}a}\right)\psi^b-(\phi\leftrightarrow\psi)=\left({\nabla}_{\alpha}{\nabla}_{\beta}+H^2{g}_{\alpha\beta}\right)\chi^{(0)a}\,,
\end{split}
\end{equation}
which must hold for arbitrary fields $h^a_{\mu\nu}$ and gauge functions $\phi$ and $\psi$.  We can eliminate the unknown function $\chi^{(0)a}$ in the above expression by operating on both sides with $\nabla_{\sigma}$ and antisymmetrizing over the indices $\sigma$ and $\alpha$ (or, equivalently, over $\sigma$ and $\beta$).  The resulting condition imposes constraints on the tensors $B^{(1)}$, $D^{(1)}$ and $C^{(1)}$, yielding the following:
\begin{equation} \label{eq:results1}
\begin{split}
\delta^{(1)}_{\phi}h_{\alpha\beta}^{a}&= \hat{d}_{1\phantom{a}bc}^{\phantom{I}a}\,F^{b}_{\lambda(\alpha\beta)}{\nabla}^{\lambda}\phi^c+\hat{d}_{2\phantom{a}bc}^{\phantom{2}a}\,{g}_{\alpha\beta}F^{b}_{\lambda}{\nabla}^{\lambda}\phi^c+\hat{d}_{3\phantom{a}bc}^{\phantom{3}a}\,F_{(\alpha}^{b}{\nabla}_{\beta)}\phi^c\\
&\quad+\hat{c}_{1\phantom{a}bc}^{\phantom{1}a}\,{\nabla}_{(\alpha}F_{\beta)}^{b}\phi^c+\hat{c}_{2\phantom{a}bc}^{\phantom{2}a}\,{\nabla}^{\lambda}F^{b}_{\lambda(\alpha\beta)}\phi^c+\hat{c}_{3\phantom{a}bc}^{\phantom{3}a}\,{g}_{\alpha\beta}{\nabla}^{\lambda}F^{b}_{\lambda}\phi^c\,.\\
\end{split}
\end{equation}
Here the hatted parameters are some linear combinations of the un-hatted ones in eq.~\eqref{eq:order1tensors}, and we are using the definitions
\begin{equation}
F_{\lambda\alpha\beta}^{a}\equiv{\nabla}_{\lambda}h_{\alpha\beta}^{a}-{\nabla}_{\alpha}h_{\lambda\beta}^{a}\,,\qquad F_{\lambda}^a\equiv {g}^{\alpha\beta}F_{\lambda\alpha\beta}^{a}\,.
\end{equation}
The tensor $F_{\lambda\alpha\beta}^{a}$ is nothing but the field strength \eqref{fieldstrengthfd} of the free PM theory, and as such it is invariant under the lowest order gauge symmetry: $\delta^{(0)}F_{\lambda\alpha\beta}^{a}=0$.
We have thus simply recovered in eq.~\eqref{eq:results1} the most general combination involving one power of $F_{\lambda\alpha\beta}^{a}$ and two derivatives, and which therefore trivially satisfies the closure condition~\eqref{eq:closure_order0} because the composition of two gauge transformations is zero. In particular, this gauge transformation is abelian, thereby ruling out a nonlinear PM theory of the Yang--Mills type.

One might ask if the obstruction could be avoided simply by using tetrads or frame fields \cite{Skvortsov:2006at,Hinterbichler:2012cn,Zinoviev:2014zka} rather than metrics since 
there are extra fields and St\"uckelberg symmetries, but as long as a metric formulation can be recovered through gauge fixing and elimination of auxiliary fields, the arguments here apply.

\section{Argument from 3-point functions}

S-matrix arguments provide strong constraints on the possible non-linear interactions of gauge theories on asymptotically flat space, {e.g.}~\cite{Weinberg:1964ew,Weinberg:1965rz,Weinberg:1980kq,Benincasa:2007xk,Schuster:2008nh,Porrati:2008rm,He:2008nj,Benincasa:2011pg,Benincasa:2011kn,Porrati:2012rd}.
The power of these arguments is that they are insensitive to field redefinition and gauge ambiguities and hence directly constrain the physical data of the theory. 

The partially massless theory lives only on de Sitter and has no flat space analogue, so we cannot directly apply standard S-matrix techniques.  However, the dual CFT correlation functions of AdS/CFT~\cite{Maldacena:1997re} and dS/CFT \cite{Strominger:2001pn} are the (A)dS analogues of the flat space S-matrix
and, just like the flat space S-matrix, their essential structure is blind to field redefinition ambiguities and gauge redundancies of the bulk theory. Any consistent field theory on AdS$_{D}$ defines a set of conformally invariant correlation functions of some effective CFT~\cite{Fitzpatrick:2010zm} on $M_d$, $d\equiv D-1$, computed using Witten diagrams~\cite{Witten:1998qj}. The goal of this section is to constrain the existence of any putative PM Yang--Mills spin-2 theory by considering the dual correlators such a theory would yield.

If a PM Yang--Mills theory of the form~\eqref{YMpa} existed, it would possess a  3-point vertex compatible with the gauge symmetry.  The possible bulk 3-point interactions for partially massless fields are studied in \cite{Joung:2012rv,Joung:2012hz}, but rather than work with them directly, we will use the fact that boundary 3-point correlators are in one-to-one correspondence with non-trivial bulk 3-point vertices~\cite{Costa:2011mg}. Given this, we can deduce whether or not a bulk vertex exists by investigating whether or not it is possible to construct the corresponding boundary correlator. We will find that for the YM-type theory we seek, the boundary 3-point correlator that would be implied by the existence of the theory does not exist, in agreement with the findings of the closure argument in Section \ref{closuresect}.

Thus we will now be considering the dual theory living on the flat-space boundary theory of AdS.\footnote{Though we are ultimately interested in the stable partially fields on dS, we nevertheless work on AdS for the dual CFT argument. The difference is simply some factors of $i$'s and minus signs -- the existence of a consistent non-linear interaction and gauge symmetry is insensitive to these differences, aside from possible imaginary coefficients in interaction terms which might rule out one case or another. Ultimately, we will find that there are no acceptable cubic interactions at all, so the argument applies immediately to dS theories as well.}  (See \cite{Bekaert:2013zya} for more on boundary values of PM fields.) The cartesian boundary coordinates are $x^i$, where $i$ ranges over the $d=D-1$ dimensions of the boundary, and the flat boundary metric is $\eta_{ij}$.  
A field in AdS$_{D}$ of spin-2 -- with usual Dirichlet boundary conditions -- corresponds to a rank-2 symmetric traceless primary operator, ${\cal O}_{ij}(x)$, of dimension $\Delta$ set by the mass of the AdS$_{D}$ field through the AdS/CFT relation 
\be 
\Delta={d\over 2}+\sqrt{{d^2\over 4}+m^2L^2}.
\ee 
Here $L^2$ is the AdS radius of curvature.
The two values of the mass with enhanced gauge symmetry are the ordinary massless graviton at $m^2=0$, and the partially massless graviton at\footnote{The conformal weight for the partially massless graviton lies below the unitarity bound $\Delta\geq d$ for spin-2 operators \cite{Mack:1975je}, reflecting the fact that the field is ghost-like on AdS.} $m^2L^2=-(d-1)$.  These correspond to dual operators which are conserved and doubly conserved, respectively:
\begin{align} 
 \partial^j {\cal O}_{ij}&=0, ~~~~~~~~\Delta=d, ~~~~~~~~~~\!~~~~~~m^2=0, \\
  \partial^i  \partial^j {\cal O}_{ij}&=0, ~~~~~~~~ \Delta=d-1, ~~~~~~~  m^2L^2=-\left(d-1\right). \label{deltapmv}
\end{align}
These conditions follow from the conformal algebra by looking for when descendants acquire zero norm~\cite{Dolan:2001ih}.  They can also be seen directly from the general form of the two-point function of two spin-2 operators of weight $\Delta$, which is fixed up to a constant by conformal symmetry \cite{Polyakov:1974gs,Osborn:1993cr,Weinberg:2010fx},
\be \langle {\cal O}_{ij}(x){\cal O}_{kl}(0)\rangle\sim {1\over x^{2\Delta}}\left(I_{ik}I_{jl}+I_{jk}I_{il}-{2\over 3}\eta_{ij}\eta_{kl}\right),\ \ \ I_{ij}\equiv\eta_{ij}-2{x_i x_j\over x^2}. \label{2points2}\ee
Taking the divergence $\partial^i$ of~\eqref{2points2}, we find a non-vanishing expression proportional to $\Delta-d$, and taking the double divergence $\partial^i\partial^j$, we find a non-vanishing expression proportional to $(\Delta-d)(\Delta-d+1)$, indicating that $\Delta=d$ is the only value for which ${\cal O}_{ij}$ can be singly conserved, and $\Delta=d-1$ is the only value for which ${\cal O}_{ij}$ can be doubly conserved but not singly conserved.

We thus turn to the construction of conformally invariant 3-point functions for these tensor operators. To catalogue the possible 3-point structures, we use the formalism of~\cite{Costa:2011mg}, which employs the embedding space formalism dating back to Dirac~\cite{Dirac:1936fq}. 
The correlators are written by contracting the fields with auxiliary vectors $z^i$ which are all null, $z^2=0$,
\be {\cal O}(x,z)={\cal O}_{ij}(x)z^iz^j .\ee
The correlation functions are obtained by stripping off the $z$'s using the operators $D_i = \left(\frac{d}{2}-1+z\cdot\frac{\partial}{\partial z}\right)\frac{\partial}{\partial z^i}-\frac{1}{2}z_{i} \frac{\partial^2}{\partial z\cdot\partial z}~,
$ which have the effect of projecting onto the symmetric traceless part of the indices contracted with the $z$'s that have been stripped,
\be \langle {\cal O}_{i_1j_1}(x_1){\cal O}_{i_2j_2}(x_2){\cal O}_{i_3j_3}(x_3)\rangle\sim D^1_{i_1} D^1_{j_1} D^2_{i_2} D^2_{j_2}D^3_{i_3} D^3_{j_3}\langle {\cal O}(x_1,z_1)  {\cal O}(x_2,z_2)  {\cal O}(x_3,z_3) \rangle \,.
\ee

The correlators are then constructed in terms of the building blocks 
\begin{align}
V_{I,JK} &\equiv   \frac{(z_J\cdot x_{IJ})\,  x_{IK}^2-(z_K\cdot x_{IK})\, x_{IJ}^2}{ x_{JK}^2},\ \ \\
H_{IJ} &\equiv   ( z_I\cdot z_J)\,  x_{IJ}^2-2(z_J\cdot x_{IJ})\, (z_I\cdot x_{IJ}), \ \ \ 
\end{align}
where $x_{IJ}^i\equiv x_I^i-x_J^i$ and $I,J,K$ range over $1,2,3$. The correlators take the explicit form
\be \langle {\cal O}^a(x_1,z_1)  {\cal O}^b(x_2,z_2)  {\cal O}^c(x_3,z_3) \rangle\sim {Q^{abc}(V,H)\over \left| x_{12}\right|^{\Delta+2}\left| x_{13}\right|^{\Delta+2}\left| x_{23}\right|^{\Delta+2}}. \label{fullcorgene}\ee
The numerator, $Q^{abc}(V,H)$, is constructed as a linear combination of color structure tensors and the following $11$ tensor structures found using the results of \cite{Costa:2011mg},
\bea && H_{12}H_{13}H_{23},\ \ \ H_{12}H_{23}V_{1,23}V_{3,21},\ \ \ H_{12}H_{13}V_{2,31}V_{3,12},\ \ \  H_{13}H_{23}V_{1,23}V_{2,13} \nonumber \\ 
&&
 H_{12}^2 V_{3,12}^2,\ \ \ H_{13}^2V_{2,31}^2,\  \ \ H_{23}^2V_{1,23}^2, \ \ \ H_{12}V_{1,23}V_{2,31}V_{3,12}^2 \nonumber \\ 
&&  H_{13}V_{1,23}V_{3,12}V_{2,31}^2,\ \ \ H_{23}V_{2,31}V_{3,12}V_{1,23}^2,\ \ \ V_{1,23}^2 V_{2,31}^2 V_{3,12}^2. \label{3point11struc}
\eea

By Bose statistics, the entire correlator must be symmetric under the simultaneous interchange of the space-time coordinates, auxiliary vectors, and color labels.  There are three possibilities for the components of the numerator: it can contain a component in the form of a totally symmetric color structure constant $f_{(S)}^{abc}$ multiplying a function $Q_{(S)}$ of the coordinate structures \eqref{3point11struc} which is totally symmetric under swapping the space-time coordinates and auxiliary vectors, $(x_I,z_I)\leftrightarrow (x_J,z_J)$,
\be 
Q^{abc}(V,H) \supset f_{(S)}^{abc}Q_{(S)}\left(x_1,z_1;x_2,z_2;x_3,z_3\right),
\ee
a totally antisymmetric color structure constant $f_{(A)}^{abc}$ multiplying a totally antisymmetric function $Q_{(A)}$ of the coordinates and auxiliary vectors,
\be 
Q^{abc}(V,H) \supset f_{(A)}^{abc}Q_{(A)}\left(x_1,z_1;x_2,z_2;x_3,z_3\right),
\ee
or a mixed symmetry color structure constant $f_{(M)}^{ab,c}$ multiplying a mixed symmetry function $Q_{(M)}$ of the coordinates and auxiliary vectors, 
\bea
&& Q^{abc}(V,H) \supset \nn\\ && f_{(M)}^{ab,c}Q_{(M)}\left(x_1,z_1;x_2,z_2 | x_3,z_3\right)+f_{(M)}^{bc,a}Q_{(M)}\left(x_2,z_2;x_3,z_3 | x_1,z_1\right)+f_{(M)}^{ca,b}Q_{(M)}\left(x_3,z_3;x_1,z_1 | x_2,z_2\right).\nn\\
\eea
Here $f_{(M)}^{ab,c}$ satisfies $f_{(M)}^{(ab),c}=0$, $f_{(M)}^{[ab,c]}=0$, and similarly for $M$ with respect to interchange of its spacetime labels.  If there is only a single field, then only the fully symmetric structure is possible.  If there are two fields, only the symmetric and mixed symmetry structures are possible.  If there are three or more fields, all are possible.  (Note that the denominator in~\eqref{fullcorgene} is totally symmetric, so we do not have to worry about its symmetry properties.)

To see which symmetry components are present, we must decompose the possible three point structures \eqref{3point11struc} according to their transformation properties under the group $S_3$ of permutations of the three labels.  
The 11-dimensional space of 3-point structures spanned by \eqref{3point11struc} transforms as a representation of the permutation group of three elements, $S_3$.
The irreducible representations of $S_3$ are the fully symmetric functions, $Q_{(S)}$, which each give a one-dimensional representation, the fully antisymmetric functions, $Q_{(A)}$, which each give a one-dimensional representation, and the mixed-symmetry functions, $Q_{(M)}$, which each give a two-dimensional representation.  

Performing this decomposition using Young projectors (see e.g. \cite{Tung:1985na}), 
we find that the 11 dimensional space of 3-point structures decomposes into 5 independent fully symmetric structures,
\be  Q_{(S)}\left(x_1,z_1;x_2,z_2 ;x_3,z_3\right) \sim \begin{cases} H_{12}H_{23}H_{13}  \\ H_{12}H_{23}V_{1,23}V_{3,12}+H_{12}H_{13}V_{2,31}V_{3,12}+H_{13}H_{23}V_{1,23}V_{2,31}\\ H_{12}^2 V_{3,12}^2+H_{13}^2 V_{2,31}^2+H_{23}^2 V_{1,23}^2 \\  H_{12}V_{1,23}V_{2,31}V_{3,12}^2+H_{13}V_{1,23}V_{3,12}V_{2,31}^2+H_{23}V_{2,31}V_{3,12}V_{1,23}^2 \\ V_{1,23}^2V_{2,31}^2V_{3,12}^2\, ,
\end{cases}
\ee

and 3 independent two-dimensional mixed symmetry representations, 
\be  Q_{(M)}\left(x_1,z_1;x_2,z_2 |x_3,z_3\right) \sim \begin{cases}  H_{13}^2 V_{2,1 3}^2 -H_{2 3}^2 V_{1,2 3}^2  \\ H_{1 2} H_{2 3} V_{1,2 3} V_{3,1 2} + 
 H_{12} H_{1 3} V_{2,1 3} V_{3,12} \\  H_{2 3} V_{1,2 3}^2 V_{2,13} V_{3,12} + 
 H_{13} V_{1,2 3} V_{2,1 3}^2 V_{3,1 2}\, .
\end{cases}
\ee

There is no fully antisymmetric representation present in the space spanned by the possible structures \eqref{3point11struc}, and therefore no possible correlator which uses an antisymmetric structure constant.  This implies that there cannot be a bulk partially massless multiplet of fields which has a non-trivial cubic vertex making use of an antisymmetric structure constant.  Insofar as this is what we mean by a partially massless Yang--Mills theory, we can now rule it out without any further calculation.

The correlators must also satisfy the double conservation conditions required for a partially massless field 
\be  \frac{\partial^2}{\partial x_1~\!\!_i\partial x_1~\!\!_j}\langle {\cal O}^a_{ij}(x_1){\cal O}^b_{kl}(x_2){\cal O}^c_{mn}(x_3)\rangle=0,  \ee
and similarly for the $x_2$, $x_3$ arguments.  This will reduce the number of allowed correlators (the number of independent correlators and the conservation condition is also affected by dimension dependent identities when $D\leq 4$ \cite{Giombi:2011rz,Costa:2011mg,Stanev:2012nq,Zhiboedov:2012bm}), however, we will not impose this condition presently because -- as we have seen -- there is no 3 point function which has the appropriate symmetries to come from a Yang--Mills like vertex in any case.

We can compare this result to the analogous result for the case of a spin-1.  For spin-1, there are 4 possible 3-point structures.  Two of them are antisymmetric, and the remaining two form a mixed symmetry structure.  Only the two antisymmetric ones are conserved for $\Delta=d-1$, and correspond to the bulk Yang--Mills coupling and to the abelian $\sim f_{abc}F_{\mu\nu}^aF^{\nu\lambda b}F_{\lambda}^{\ \mu c}$ coupling.  The fact that there are no fully symmetric structures tells us that there is no non-trivial cubic self-coupling for a single photon\footnote{In quantum electrodynamics, this is also a consequence of Furry's theorem \cite{Furry:1937zz}.}.

\section{Conclusions}
We have argued that there does not exist a theory which might reasonably be described as a multiplet of partially massless spin-2 fields interacting in a Yang--Mills like fashion. We gave two lines of evidence for this statement.  The first comes from considering directly the gauge symmetries that such a theory might possess; we have explicitly checked that any putative deformation of the gauge symmetry which is linear in the fields (as in the case of Yang--Mills) only closes to form an algebra if it is abelian. The second comes from considering the dual correlation functions which a PM Yang--Mills theory would give through AdS/CFT.  We expect that any theory defined on (A)dS space should define correlation functions for an effective conformal field theory defined on the boundary, and in the case of a PM Yang--Mills theory, the 3 point boundary correlator should take the form of an antisymmetric structure constant times an antisymmetric function of the spacetime coordinates, as it does in regular Yang--Mills theory.  However, we have seen that no such structures exist for a PM spin-2 which are consistent with conformal invariance.

Our no-go arguments do not rule out {\it any} theory of interacting PM spin-2 fields, but they do strongly constrain such a theory.
Our two arguments suggest that if a theory of multiple interacting partially massless spin-2 fields does exist, it will be quite different from a Yang--Mills type theory.  The arguments suggest where possible loopholes might be found. One loophole in our arguments is that we have considered only the structure of the cubic vertices of any putative theory; it is possible that an interacting theory may only start at quartic or higher order, corresponding to a zero 3 point function and a deformation of the gauge symmetry which begins at higher order in the fields.  Another is that in the closure argument we have restricted to two-derivative extensions of the gauge symmetry, it is possible that allowing for more derivatives could change the story, however to be consistent with the correlator arguments any cubic vertices would again have to be on-shell trivial.  We do find that there are mixed symmetry 3 point structures which one might imagine could form the basis of a non-abelian theory, but in this case again the closure arguments seem to preclude these vertices deforming the algebra.  Finally, it could be that the partially massless spin-2 has more in common with higher spin gauge fields than it does with gravity or electromagnetism; it is widely believed that the only way to construct consistent theories of higher spin gauge theories is to include an infinite number of them, as in Vasiliev theory \cite{Vasiliev:1990en}, and the same may be true for the partially massless spin-2.

\noindent
{\bf Acknowledgements:}  KH, AJ and RR would like to thank the Sitka Sound Science Center for their hospitality while some of this work was completed. Research at Perimeter Institute is supported by the Government of Canada through Industry Canada and by the Province of Ontario through the Ministry of Economic Development and Innovation.  This work was made possible in part through the support of a grant from the John Templeton Foundation. The opinions expressed in this publication are those of the author and do not necessarily reflect the views of the John Templeton Foundation (KH). This work was supported in part by the Kavli Institute for Cosmological Physics at the University of Chicago through grant NSF PHY-1125897, an endowment from the Kavli Foundation and its founder Fred Kavli, and by the Robert R. McCormick Postdoctoral Fellowship (AJ).  EM is supported by the Swiss National Science Foundation.  RAR is supported by DOE grant DE-SC0011941.  We make use of \cite{citeulike:13127953} for some of the calculations in Section \ref{closuresect}.

\renewcommand{\em}{}
\bibliographystyle{utphys}
\addcontentsline{toc}{section}{References}
\bibliography{PM-yang-mills-resubmit}

\end{document}